\documentclass[12pt,aps,showpacs,eqsecnum,nofootinbib,floatfix]{revtex4}
\usepackage{bbding}
\usepackage{pifont}
\usepackage{subfigure}
\usepackage{epsfig}
\usepackage{array}
\usepackage{amsmath}
\usepackage{amsfonts}
\usepackage{amssymb}
\usepackage{color}
\usepackage{graphicx}
\usepackage{mathrsfs}
\usepackage{multirow}
\usepackage{subfigure}
\usepackage[colorlinks,citecolor=blue, linkcolor=blue,hyperindex,dvipdfm]{hyperref}

\def\be{\begin{equation}}
\def\ee{\end{equation}}
\def\bea{\begin{eqnarray}}
\def\eea{\end{eqnarray}}

\newcommand{\omits}[1]{}

\begin{document}

\title{Entropy of higher dimensional topological dS black holes with
nonlinear source}
\author{Yang Zhang$^{a,b}$, Li-Chun Zhang$^{a,b} $ and Ren Zhao$^{b}$}
\affiliation{{\footnotesize $^a$Department of Physics, Shanxi Datong University, Datong
037009, China}\\
{\footnotesize $^b$Institute of Theoretical Physics, Shanxi Datong
University, Datong 037009, China}}

\begin{abstract}
On the basis of the first law of black hole thermodynamics, we propose the
concept of effective temperature of de Sitter(dS) black holes and conjecture
that the effective temperature should be the temperature of the dS black
holes when the Hawking radiation temperatures of the black hole horizon and
the cosmological horizon are equal. Choosing different independent
variables, we can find a differential equation satisfied by the entropy of
the dS black hole. It is shown that the differential equation of entropy is
independent of the choice of independent variables. From the differential
equation, we get the entropy of dS black hole and other effective
thermodynamic quantities. We also discuss the influence of several
parameters on the effective thermodynamic quantities.
\end{abstract}

\pacs{04.70.-s, 05.70.Ce}
\maketitle

\section{Introduction}

Recent years, the studies on the thermodynamic properties of black holes
have made some progress. Firstly, Based on the fact that black hole
satisfies the first law of thermodynamics, one corresponded the cosmological
constant in AdS spacetime to the pressure in general thermodynamics, and got
the extended first law of thermodynamics of black holes. Various critical
phenomena of black holes in AdS spacetime have been extensively studied by
comparing the equation of state of black hole with the van der Waals
equation . One can further study the critical point and critical index of
the AdS black holes, and discussed the influence of the parameters in
spacetime on the phase transition \cite%
{1,2,3,4,5,6,7,8,9,10,11,12,13,14,15,16,17,18,19,20,21,22,23,24,25,26}.

For de Sitter spacetime, thermodynamic quantities of black hole horizon and
cosmological horizon satisfies with the first law of thermodynamics
respectively \cite{27,28}. However, the Hawking radiation temperature of two
horizons are not equal in general cases \cite{+1,+2,+3,+4,+7,29,30,31,32},
thus the de Sitter black holes does not meet the requirement of stability of
the thermodynamic equilibrium. This limited the research on the
thermodynamical properties of the de Sitter black holes. Recent years, with
the in-depth study of dark energy, the thermodynamical properties of de
Sitter black holes aroused people's attention \cite{35,36,37,38,39,40}.
Because during the inflation in the very early period, our Universe was a
quasi-de Sitter spacetime, and the cosmological constant introduced in the
de Sitter spacetime is the contribution of vacuum energy, which is also a
kind of matter energy. If the cosmological constant is just the dark energy,
our Universe will evolve to a new de Sitter phase \cite{+8,+9}. For
constructing the whole history of evolution of the Universe, it's necessary
to understand both the classical and quantum properties of the de Sitter
spacetime. People usually define the entropy of de Sitter spacetime as the
sum of two horizon's entropies \cite{29,32,35,36,37}. However, there is no
theoretical proof supports this consequence.

Reviewing the four laws of thermodynamics found by Hawking and Bekenstein
etc., \cite{41,42,43,44,45}, and given the Hawking radiation temperature of
black hole, the equation of state of black hole satisfy the first law of
thermodynamics. By comparing the such a equation with the general
thermodynamics system, they got the famous conclusion that the entropy of
black hole equals quarter area of black hole horizon. Using this approach to
the de Sitter spacetime one can get the same conclusion, i.e. the entropy of
black hole horizon and the cosmological horizon are quarter of their
corresponding area respectively. But some thermodynamic quantities in two
horizons are same, such as energy, charge etc., which means two horizons are
not independent. Entropy is usually related to the numbers of microscopic
states. Although the microscopic origin of black hole entropy is still
unclear, it must exist. For black holes with multiple horizons, such as SdS
black hole, the black hole horizon and the cosmological horizon are in fact
not independent. There may exist some correlations between them. The size of
black hole horizon is closely related to the size of the cosmological
horizon, and the evolution of black hole horizon will lead to the evolution
of the cosmological horizon. Taking the correlations between the horizons
into account, the total numbers of microscopic states are not simply the
product of those of the two horizons (it will be, if the two horizons are
isolated). Therefore, the total entropy is no longer the sum of the
entropies of the two horizons, but should include an extra term coming from
the contribution of the correlations of the two horizons.

In order to further study the thermodynamic properties of de Sitter black
holes, we need to define its temperature and entropy. In this work, based on
the consideration of dimension, we set the entropy of de Sitter black holes
has the form of function $F_{n}(x)$. (here $x=r_{+}/r_{c}$ and $r_{+},r_{c}$
are positions of the black hole horizon and the cosmological horizon).
Through the basic condition that the thermodynamic quantities of spacetime
satisfy the first law of thermodynamics, and the relation between the
effective temperature of spacetime and the radiation temperature of two
horizons, we can find the differential equation for $F_{n}(x)$. As the black
hole horizon of spacetime approaches to zero and de Sitter black hole
approaches to the pure de Sitter one, and regard these as the initial
conditions, the differential equation can be solved and the result is just
the entropy. Furthermore, we can get the effective temperature and pressure
etc. In order to make the discussion more general, we focus on the higher
dimensional topological de Sitter black holes with nonlinear source (HDBRN)
spacetime.

This paper is organized as follows. In Sect. 2, the corresponding
thermodynamic quantities of black hole horizon and the cosmological horizon
in the HDBRN spacetime will be introduced. And the condition obeyed by
spacetime charge $Q$, when the radiation temperature of two horizons equal
to each other. In Sect. 3, on the basis of considering the relation between
two horizons, we give the entropy of HDBRN spacetime which satisfies the
first law of thermodynamics as well as the effective temperature and
pressure. Our conclusions are presented in Sect. 4.(we use the units $%
G=\hbar =k_{B}=c=1$)

\section{Topological black hole with nonlinear source}

The$(n+1)-$dimensional action of Einstein gravity of nonlinear
electrodynamics is \cite{47,48}:
\begin{equation}
{I_{G}}=-\frac{1}{{16\pi }}\int_{M}{{d^{n+1}}x\sqrt{-g}[R-2\Lambda +L(F)]}-%
\frac{1}{{8\pi }}\int_{\partial M}{{d^{n}}x\sqrt{-\gamma }}\Theta (\gamma )
\label{2.1}
\end{equation}%
where $R$ is the scalar curvature, $\Lambda $ is the cosmological constant.
In this action,
\begin{equation}
{L}(F)=-F+\alpha {F^{2}}+o({\alpha ^{2}}),  \label{2.2}
\end{equation}%
is the Lagrangian of nonlinear electrodynamics. $F={F_{\mu \nu }}{F^{\mu \nu
}}$ is the Maxwell invariant, in which ${F_{\mu \nu }}={\partial _{\mu }}{%
A_{\nu }}-{\partial _{\nu }}{A_{\mu }}$ is the electromagnetic field tensor
and $A_{\nu }$ is the gauge potential. In addition, $\alpha $ denotes
nonlinearity parameter which is small, so the effects of nonlinearity can be
considered as a perturbation.

The $n+1$ -dimensional topological black hole solutions can take the form of
\begin{equation}
ds^{2}=-f(r)d{t^{2}}+\frac{{d{r^{2}}}}{{f(r)}}+{r^{2}}d\Omega _{n-1}^{2},
\label{2.3}
\end{equation}%
where
\begin{equation}
f(r)=k-\frac{m}{{{r^{n-2}}}}-\frac{{2\Lambda {r^{2}}}}{{n(n-1)}}+\frac{{2{%
q^{2}}}}{{(n-1)(n-2){r^{2n-4}}}}-\frac{{4{q^{4}}\alpha }}{{(3{n^{7}}-7n+4){%
r^{2n-6}}}},  \label{2.4}
\end{equation}%
m is an integration constant which is related to the mass of the black hole
and the last term of Eq. (\ref{2.4}) indicates the effect of nonlinearity.
The asymptotical behavior of the solution is AdS or dS provided $\Lambda <0$
or $\Lambda >0$, and the case of asymptotically flat solution is permitted
for $\Lambda =0$ and $k=1$.

When $\Lambda >0$, the position of the spacetime black hole horizon $r_{+}$
and the universal horizon $r_{c}$ is satisfied with the equation $f({r_{+,c}}%
)=0$. The radiation temperature of the two horizon can be written as follows
\cite{28,47}
\begin{equation}
{T_{+}}=\frac{{f^{\prime }({r_{+}})}}{{4\pi }}=\frac{1}{{2\pi (n-1)}}\left( {%
\frac{{(n-1)(n-2)k}}{{2{r_{+}}}}-\Lambda {r_{+}}-\frac{{{q^{2}}}}{{%
r_{+}^{2n-3}}}+\frac{{2{q^{4}}\alpha }}{{r_{+}^{4n-5}}}}\right) .
\label{2.5}
\end{equation}%
\begin{equation}
{T_{c}}=-\frac{{f^{\prime }({r_{c}})}}{{4\pi }}=-\frac{1}{{2\pi (n-1)}}%
\left( {\frac{{(n-1)(n-2)k}}{{2{r_{c}}}}-\Lambda {r_{c}}-\frac{{{q^{2}}}}{{%
r_{c}^{2n-3}}}+\frac{{2{q^{4}}\alpha }}{{r_{c}^{4n-5}}}}\right) .
\label{2.6}
\end{equation}%
The mass of the black hole is
\begin{equation}
M=\frac{{{V_{n-1}}(n-1)}}{{16\pi }}\left(
\begin{array}{c}
{kr_{+}^{n-2}-\frac{{2\Lambda r_{+}^{n}}}{{n(n-1)}}+\frac{{2{q^{2}}}}{{%
(n-1)(n-2)r_{+}^{n-2}}}-} \\
{\frac{{4{q^{4}}\alpha }}{{[2(n-2)(n+2)+(n-3)(n-4)]r_{+}^{3n-4}}}}%
\end{array}%
\right) ,  \label{2.7}
\end{equation}%
or
\begin{equation}
M=\frac{{{V_{n-1}}(n-1)}}{{16\pi }}\left(
\begin{array}{c}
{kr_{c}^{n-2}-\frac{{2\Lambda r_{c}^{n}}}{{n(n-1)}}+\frac{{2{q^{2}}}}{{%
(n-1)(n-2)r_{c}^{n-2}}}-} \\
{\frac{{4{q^{4}}\alpha }}{{[2(n-2)(n+2)+(n-3)(n-4)]r_{c}^{3n-4}}}}%
\end{array}%
\right) ,  \label{2.8}
\end{equation}%
where ${V_{n-1}}:=\frac{{2{\pi ^{n/2}}}}{{\Gamma (n/2)}},{S_{+,c}}:=\frac{{{%
V_{n-1}}r_{+,c}^{n-1}}}{4},{V_{+,c}}:=\frac{{{V_{n-1}}r_{+,c}^{n}}}{n},Q:=%
\frac{q}{{4\pi }}{V_{n-1}},{\Phi _{+,c}}:=\frac{q}{{(n-2)r_{+,c}^{n-2}}}-%
\frac{{4{q^{3}}\alpha }}{{(3n-4)r_{+,c}^{3n-4}}},P:=-\frac{\Lambda }{{8\pi }}%
,M:=\frac{{{V_{n-1}}(n-1)m}}{{16\pi }}.$

The thermodynamic quantity corresponding to the two horizons satisfies the
first law of thermodynamics
\begin{equation}
dM={T_{+,c}}d{S_{+,c}}+{\Phi _{+,c}}dQ+{V_{+,c}}dP.  \label{2.9}
\end{equation}%
From the equation $f({r_{+,c}})=0,$ one can obtain
\begin{eqnarray}
\frac{{2\Lambda }}{{n(n-1)}} &=&\frac{{k(1-{x^{n-2}})}}{{r_{c}^{2}(1-{x^{n}})%
}}-\frac{{2{q^{2}}(1-{x^{n-2}})}}{{(n-1)(n-2)r_{c}^{2n-2}{x^{n-2}}(1-{x^{n}})%
}}  \label{2.10} \\
&&+{\frac{{4{q^{4}}\alpha (1-{x^{3n-4}})}}{{%
[2(n-2)(n+2)+(n-3)(n-4)]r_{c}^{4n-4}{x^{3n-4}}(1-{x^{n}})}}},  \notag
\end{eqnarray}%
\begin{eqnarray}
M &=&\frac{{{V_{n-1}}(n-1)}}{{16\pi (1-{x^{n}})}}r_{c}^{n-2}{k({x^{n-2}}-{%
x^{n}})}+\frac{{2{q^{2}}}}{{(n-1)(n-2)r_{c}^{2(n-2)}}}\frac{{(1-{x^{2n-2}})}%
}{{{x^{n-2}}}}  \notag \\
&&{-\frac{{4{q^{4}}\alpha }}{{[2(n-2)(n+2)+(n-3)(n-4)]r_{c}^{2(2n-3)}}}\frac{%
{(1-{x^{4n-4}})}}{{{x^{3n-4}}}}},  \label{2.11}
\end{eqnarray}%
where $x=r_{+}/r_{c}$. When $T_{+}=T_{c}$, by solving Eq (\ref{2.5}) and Eq (%
\ref{2.6}), we can get
\begin{equation}
\Lambda =\frac{{(n-1)(n-2)k}}{{2r_{c}^{2}x}}-\frac{{{q^{2}}}}{{r_{c}^{2n-2}{%
x^{2n-3}}}}\frac{{1+{x^{2n-3}}}}{{(1+x)}}+\frac{{2{q^{4}}\alpha }}{{%
r_{c}^{4n-4}{x^{4n-5}}}}\frac{{1+{x^{4n-5}}}}{{(1+x)}}.  \label{2.12}
\end{equation}%
From Eq \ref{2.10}) and Eq (\ref{2.12}), one can find when $T_{+}=T_{c}$,
one can obtain the following equation
\begin{eqnarray}
&&{l}\frac{{4{q^{4}}\alpha }}{{r_{c}^{4n-5}}}(n-2)\frac{{(3{n^{2}}-7n+4)(1-{%
x^{n}})(1+{x^{4n-5}})-n(n-1){x^{n-1}}(1-{x^{3n-4}})(1+x)}}{{(3{n^{2}}-7n+4){%
x^{2n-2}}}}  \notag \\
&=&\frac{{k(n-1)(n-2)(1+x){x^{2n-4}}}}{{{r_{c}}}}[nx(1-{x^{n-2}})-(n-2)(1-{%
x^{n}})]  \notag \\
&&-\frac{{2{q^{2}}}}{{r_{c}^{2n-3}}}[n{x^{n-1}}(1+x)(1-{x^{n-2}})-(n-2)(1-{%
x^{n}})(1+{x^{2n-3}})].  \label{2.13}
\end{eqnarray}%
The Eq. (\ref{2.13}) can be rewritten as
\begin{equation}
\frac{{4{q^{4}}\alpha }}{{r_{c}^{4n-6}}}{B_{1}}(x,n)=k{A_{1}}(x,n)+\frac{{2{%
q^{2}}}}{{r_{c}^{2n-4}}}{C_{1}}(x,n).  \label{2.14}
\end{equation}%
We can find that Eq. (\ref{2.14}) is a function relation between the $\frac{{%
4{q^{4}}\alpha }}{{r_{c}^{4n-6}}}$, $\frac{{2{q^{2}}}}{{r_{c}^{2n-4}}}$and $%
k $, and any one of these can be a function of other two independent
variables. When regard $\frac{{2{q^{2}}}}{{r_{c}^{2n-4}}}$as a function of $%
\frac{{4{q^{4}}\alpha }}{{r_{c}^{4n-6}}}$and $k$ , and substituted Eq. (\ref%
{2.12}) and Eq. (\ref{2.13}) into Eq. (\ref{2.5}) or Eq. (\ref{2.6}), thus
when the two horizon radiation temperature satisfied the relation $%
T_{+}=T_{c}$, one can obtain
\begin{eqnarray}
T &=&{T_{c}}={T_{+}}  \notag \\
&=&\frac{1}{{2\pi (n-1){r_{c}}K(x,n)}}\{k(n-2)(n-1)[n{x^{n-2}}(1-{x^{2}})(1-{%
x^{n-2}})-(1-{x^{n}})(1-{x^{2n-4}})]  \notag \\
&&+\frac{{(1-{x^{2n-2}})}}{{(3{n^{2}}-7n+4){x^{3n-4}}}}\frac{{4{q^{4}}\alpha
}}{{r_{c}^{4n-6}}}[4{x^{n-2}}(1-{x^{n}})+n(1-11{x^{n-2}}+11{x^{2n-2}}-{%
x^{3n-4}})  \notag \\
&&-2{n^{2}}(1-{x^{3n-4}}+5{x^{2n-2}}-5{x^{n-2}})+{n^{3}}(1-{x^{3n-4}}-3{%
x^{n-2}}+3{x^{2n-2}})]\},  \label{2.15}
\end{eqnarray}%
with
\begin{equation}
K(x,n)=[n{x^{n-1}}(1+x)(1-{x^{n-2}})-(n-2)(1-{x^{n}})(1+{x^{2n-3}})].
\label{2.16}
\end{equation}

When set $n=3,\alpha =0,$ the spacetime will return to RN-de Sitter
spacetime, one can obtain the effective temperature
\begin{eqnarray}
T &=&{T_{c}}={T_{+}=}\frac{k}{2\pi r_{c}}\frac{%
3x(1-x^{2})(1-x)-(1-x^{3})(1-x^{2})}{3x^{2}(1+x)(1-x)-(1-x^{3})(1+x^{3})}
\label{2.17} \\
&=&\frac{k}{2\pi r_{c}}\frac{1-x}{(1+x)^{2}},  \notag
\end{eqnarray}%
which is as same as the result mentioned by Ref \cite{30,31,cai}.

When $\frac{{4{q^{4}}\alpha }}{{r_{c}^{4n-6}}}$ is regarded as the function
of $\frac{{2{q^{2}}}}{{r_{c}^{2n-4}}}$and $k$, in the same way as above
mentioned, we can get the following equation
\begin{eqnarray}
T &=&{T_{c}}={T_{+}}=\frac{1}{{2\pi (n-1){r_{c}}(3{n^{2}}-7n+4)\tilde{K}(x,n)%
}}{(n-1)(n-2)k}  \notag \\
&&{4(1-{x^{n}})(1-{x^{4n-6}})}+n[-7(1+{x^{5n-6}})-3{x^{n-2}}(1+{x^{3n-2}})+10%
{x^{n}}(1+{x^{3n-6}})]  \notag \\
&&+{n^{2}}[5{x^{n-2}}(1+{x^{3n-2}})-8{x^{n}}(1+{x^{3n-6}})+3(1+{x^{5n-6}})]+2%
{n^{3}}[{x^{n}}(1+{x^{3n-6}})\left. {-{x^{n-2}}(1+{x^{3n-2}})]}\right]
\notag \\
&&-\frac{{2{q^{2}}(1-{x^{2n-2}})}}{{r_{c}^{2n-4}{x^{n-2}}}}{4{x^{n-2}}(1-{%
x^{n}})}+n(1-11{x^{n-2}}+11{x^{2n-2}}-{x^{3n-4}})  \notag \\
&&-2{n^{2}}(1-{x^{3n-4}}+5{x^{2n-2}}-5{x^{n-2}})+{n^{3}{(1-{x^{3n-4}}-3{%
x^{n-2}}+3{x^{2n-2}})}},  \label{2.18}
\end{eqnarray}%
where
\begin{equation}
\tilde{K}(x,n)=(n-2)\frac{{(3{n^{2}}-7n+4)(1-{x^{n}})(1+{x^{4n-5}})-n(n-1){%
x^{n-1}}(1-{x^{3n-4}})(1+x)}}{{(3{n^{2}}-7n+4)}}.  \label{2.19}
\end{equation}%
\begin{figure}[tbp]
\centering
\subfigure[]{\includegraphics[width=0.4\columnwidth,height=1.2in]{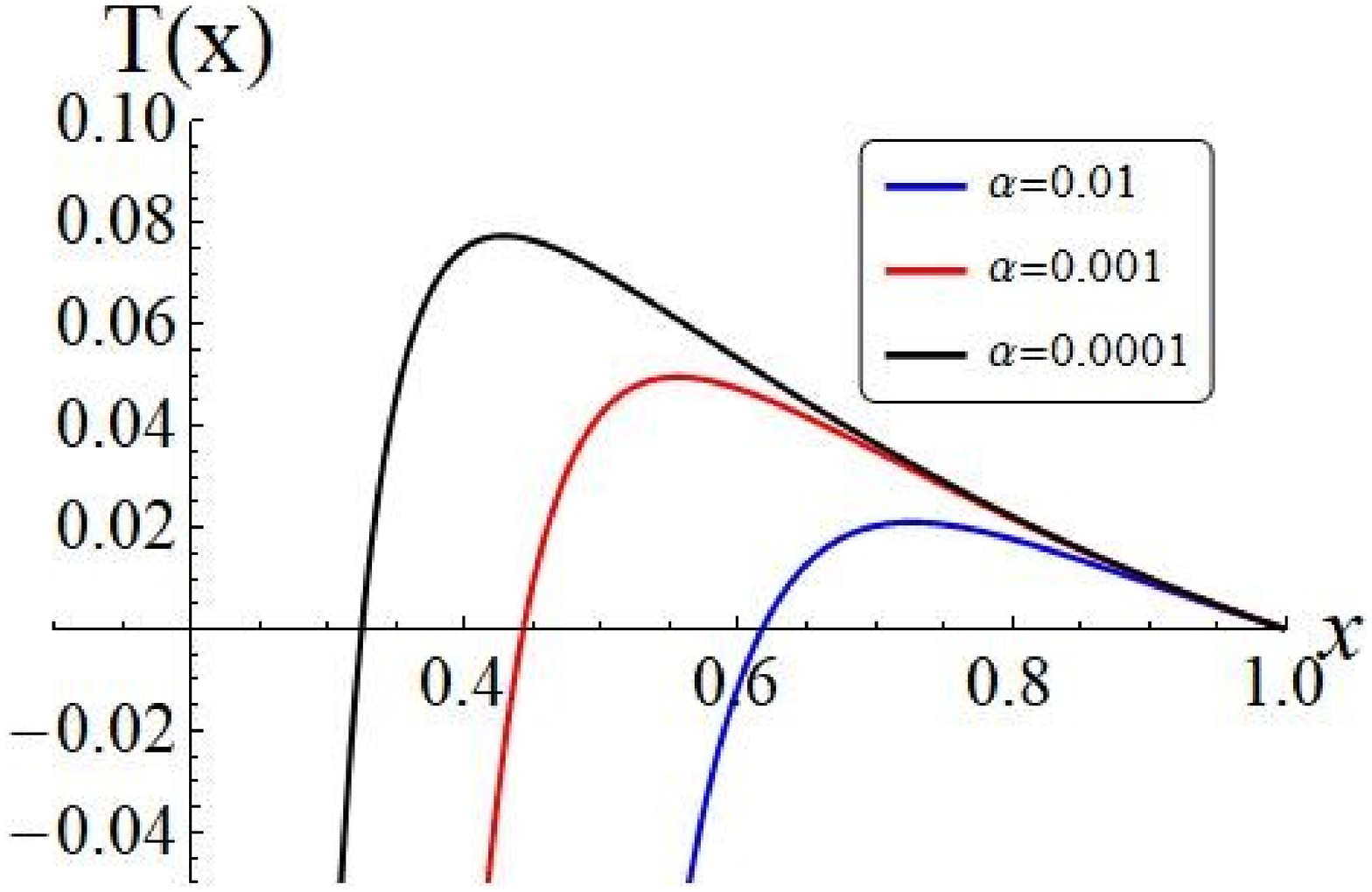}%
\label{a}}
\subfigure[]{\includegraphics[width=0.4\columnwidth,height=1.2in]{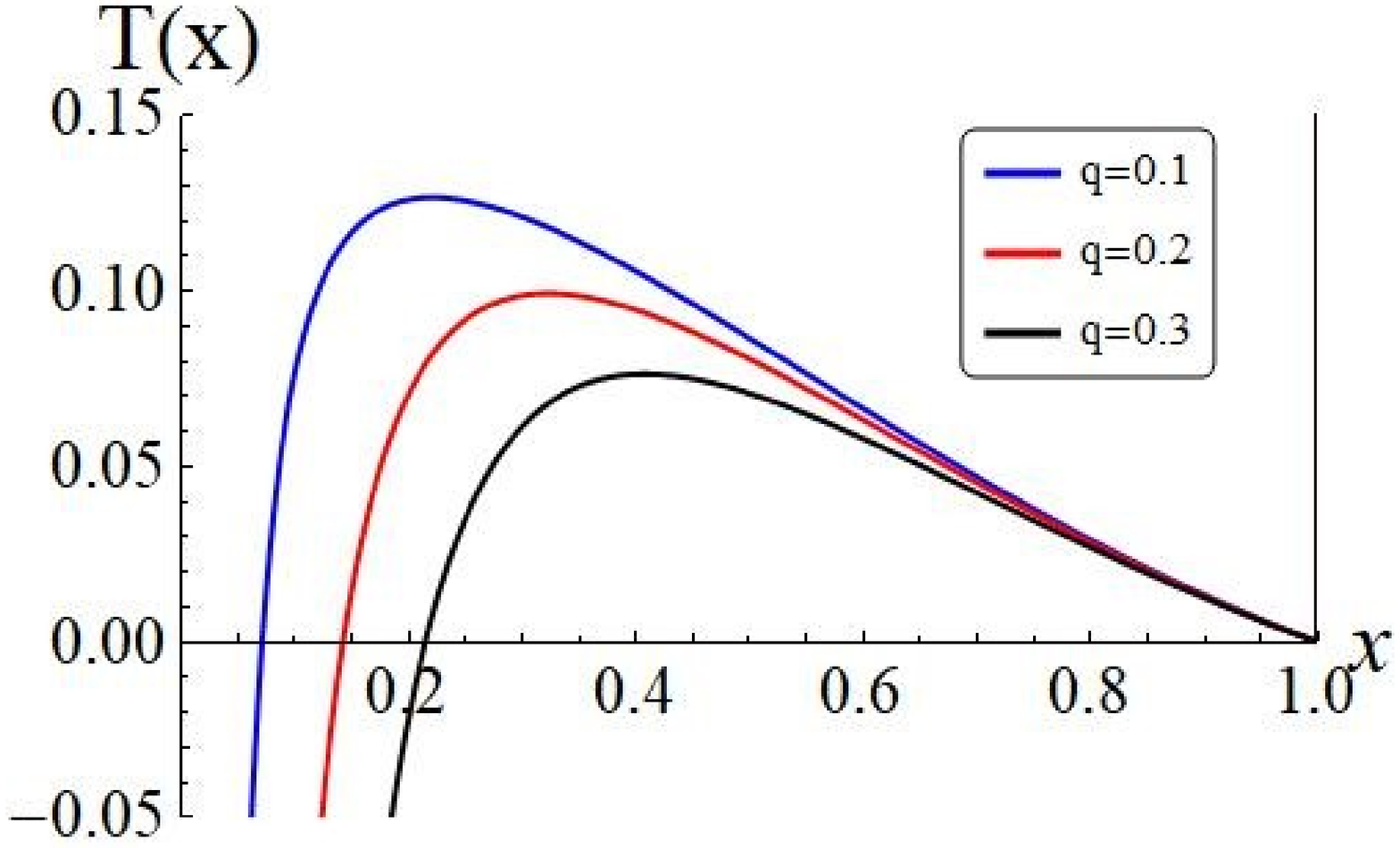}\label{b}
}
\caption{(color online).(a): The radiation temperature $T$ as a function of $%
x$. The blue line is for $\protect\alpha \mathrm{\ =}0.0001$, whereas the
red curve stands for $\protect\alpha \mathrm{\ =}0.01$. Furthermore, the
black line is for $\protect\alpha \mathrm{\ =}0.001$. Other parameters are
as follows:$r_{c}=1,n=4,q=0.1$. (b): The radiation temperature $T$ as Versus
the black hole horizon $x$ with the different values of $q$. The blue line
is for $q\mathrm{\ =}0.1$, whereas the red curve stands for $\protect\alpha
\mathrm{\ =}0.2$. Furthermore, the black line is for $\protect\alpha \mathrm{%
\ =}0.3$. Other parameters are as follows:$r_{c}=1,n=4,\protect\alpha %
=0.0001 $. }
\end{figure}

In Fig. 1(a), we plot the radiation temperature $T$ as the function of
positions of the black hole horizon $x$ with the different values of $\alpha$%
. As shown in the figure, one can see that the maximal value of the two
horizon radiation temperature is proportional to $\alpha$, and the region
where the temperature is greater than zero is also proportional to the value
of $\alpha$ . In Fig.1 (b), we also plot the radiation temperature $T$ as
the function of positions of the black hole horizon $x$ with the different
values of $q$, we find that the maximal value of the two horizon radiation
temperature is proportional to $q$ when choose the same values of $\alpha$,
and the region where the temperature is greater than zero is also
proportional to the value of $q$.

\section{Space-time entropy and the effective thermodynamic quantities}

\begin{figure}[tbp]
\centering
\subfigure[]{\includegraphics[width=0.4\columnwidth,height=1.2in]{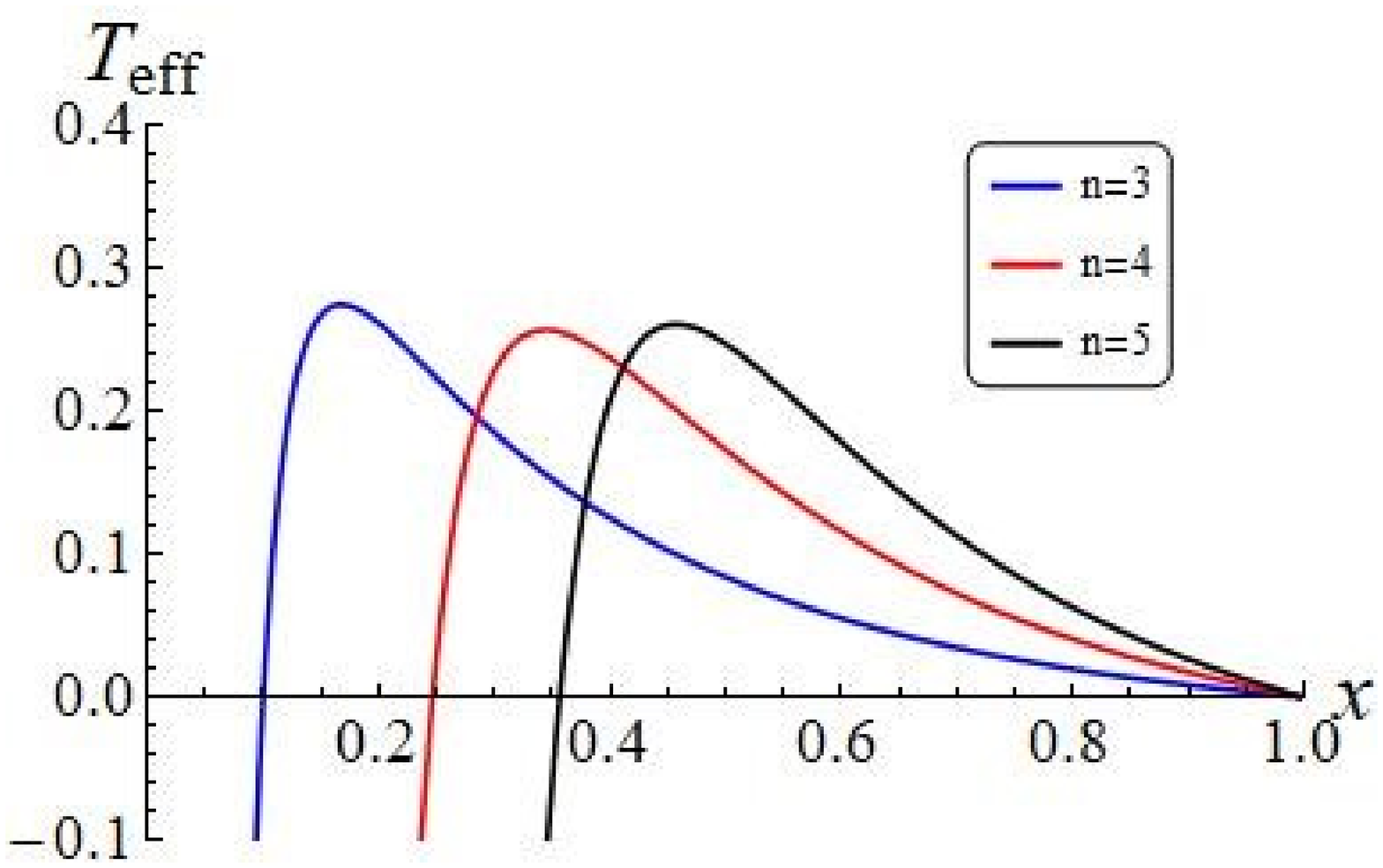}%
\label{a}}
\subfigure[]{\includegraphics[width=0.4\columnwidth,height=1.2in]{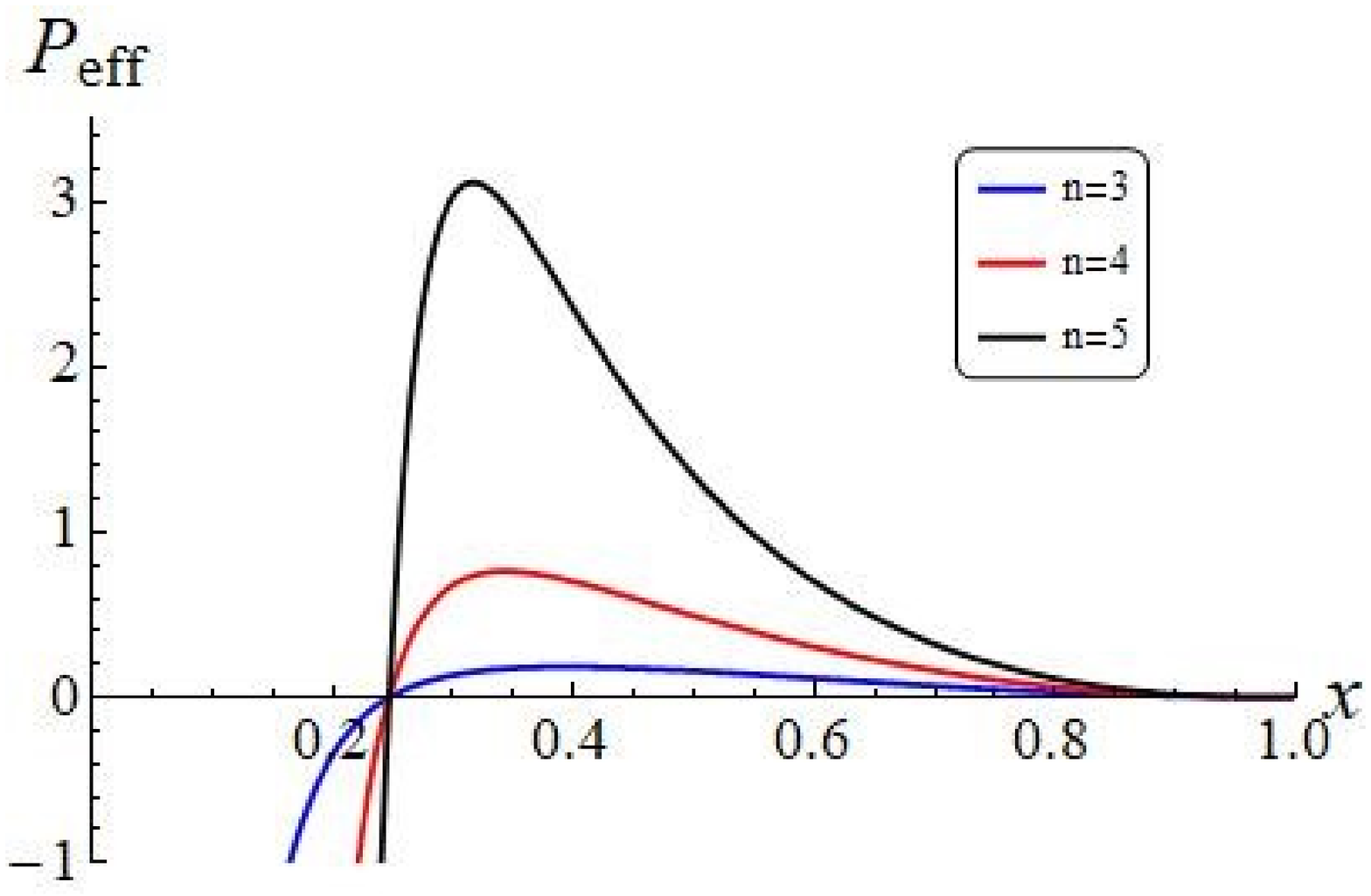}\label{b}
}\subfigure[]{\includegraphics[width=0.4%
\columnwidth,height=1.2in]{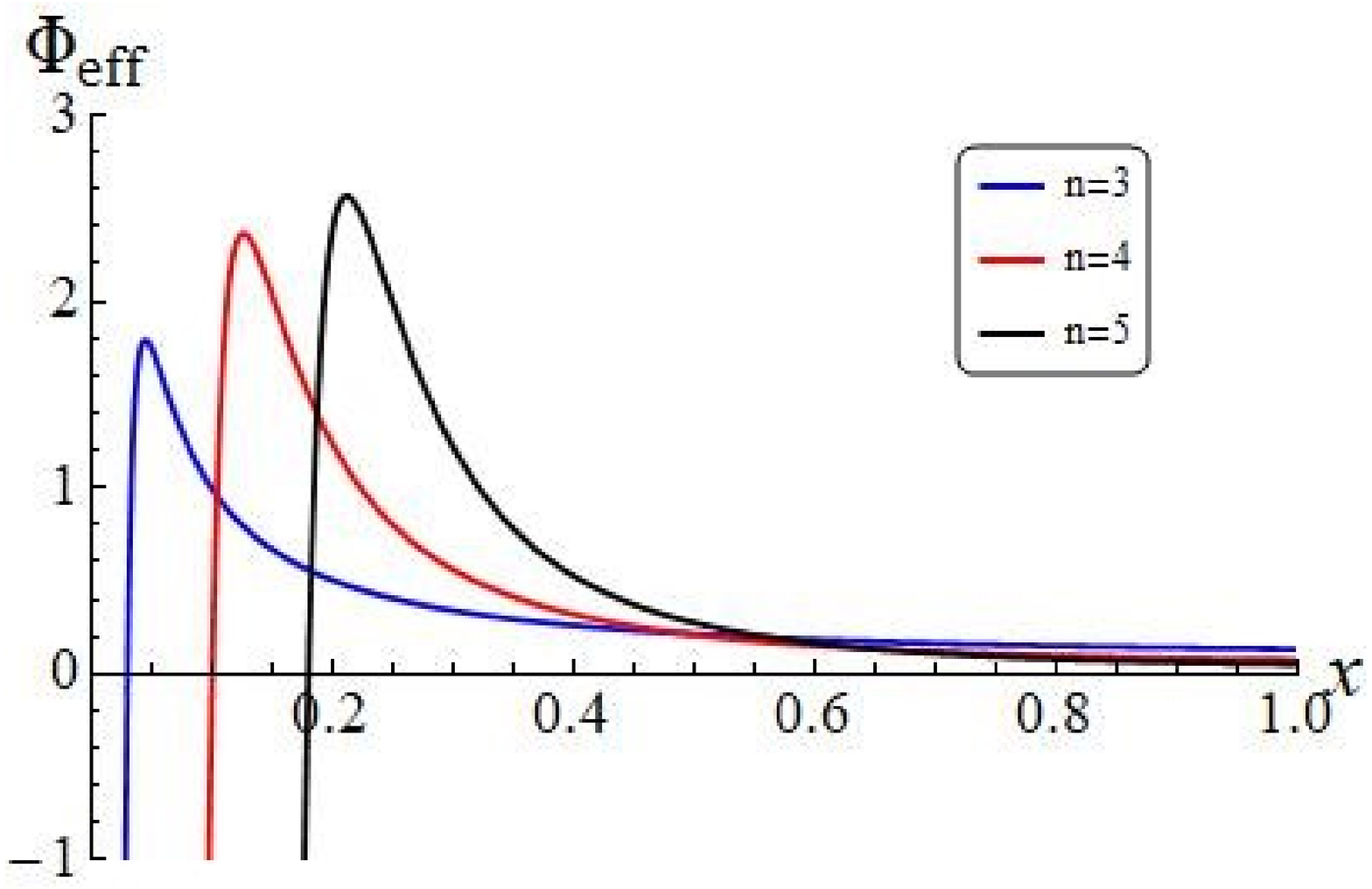}\label{c}}\newline
\caption{(color online).(a): The effective temperature $T_{eff}$ versus $x$
for the different values of spacetime dimension $n$ : $n=3$ (blue line), $%
n=4 $ (red line),$n=5$ (black line). (b): The effective pressure $p_{eff}$
versus $x$ for the different values of spacetime dimension $n$ : $n=3$ (blue
line), $n=4$ (red line),$n=5$ (black line). (c): The effective electric
potential $\Phi _{eff}$ versus $x$ for the different values of spacetime
dimension $n$ : $n=3$ (blue line), $n=4$ (red line),$n=5$ (black line).
Here, $q=0.1,\protect\alpha =0.0001$. }
\end{figure}
\begin{figure}[tbp]
\centering
\subfigure[]{\includegraphics[width=0.4\columnwidth,height=1.2in]{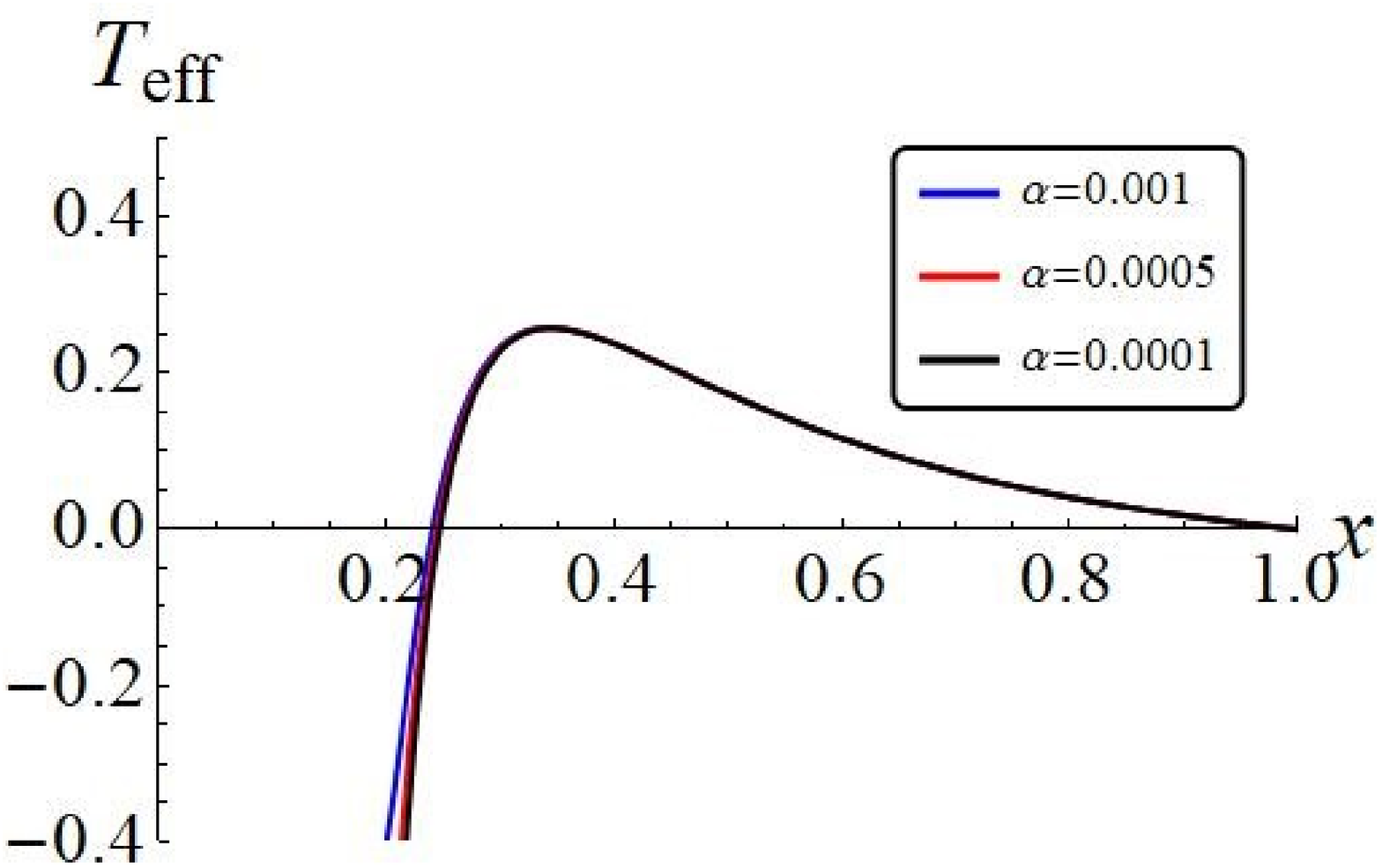}%
\label{a}}
\subfigure[]{\includegraphics[width=0.4\columnwidth,height=1.2in]{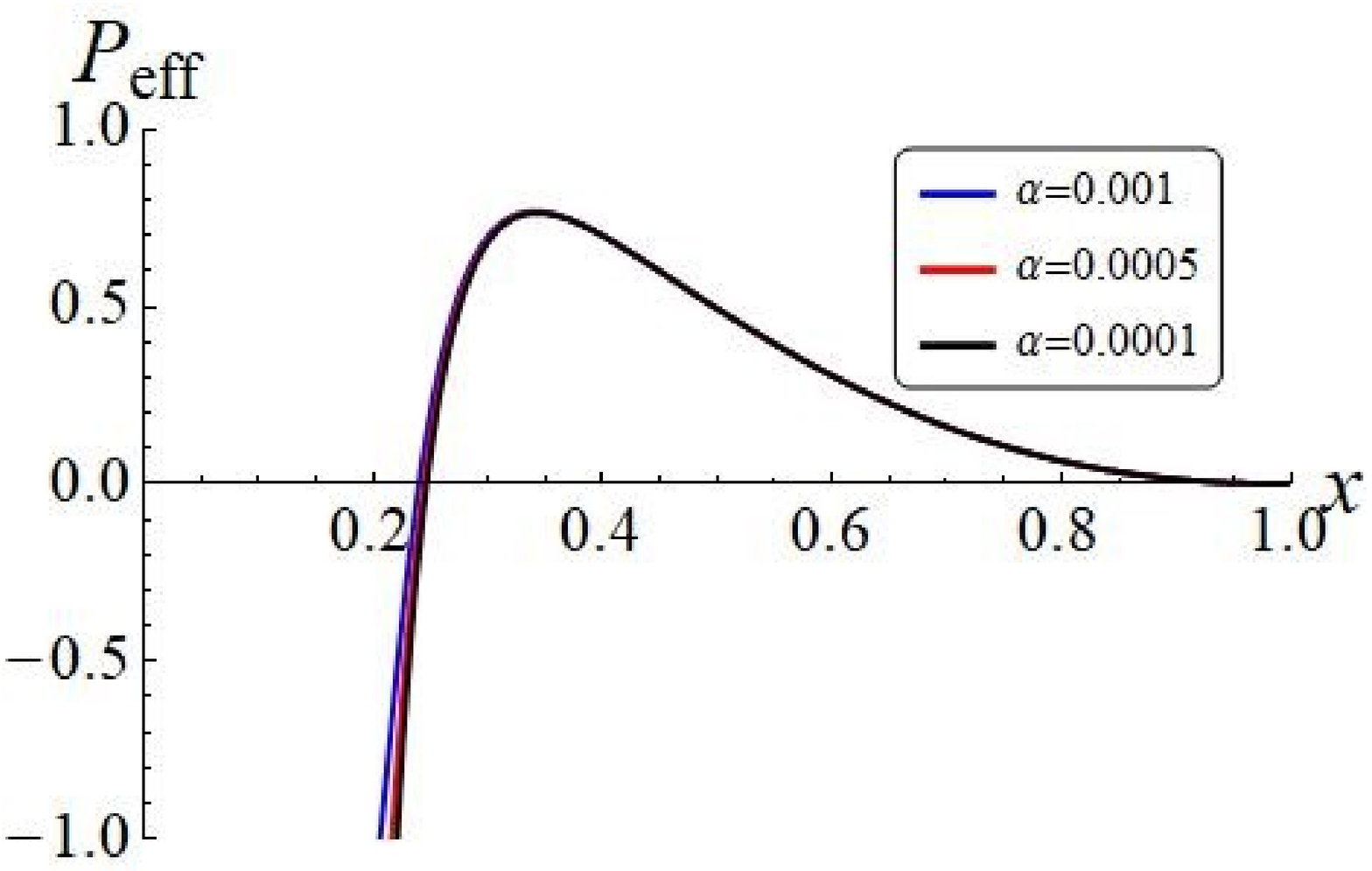}\label{b}
}\subfigure[]{\includegraphics[width=0.4%
\columnwidth,height=1.2in]{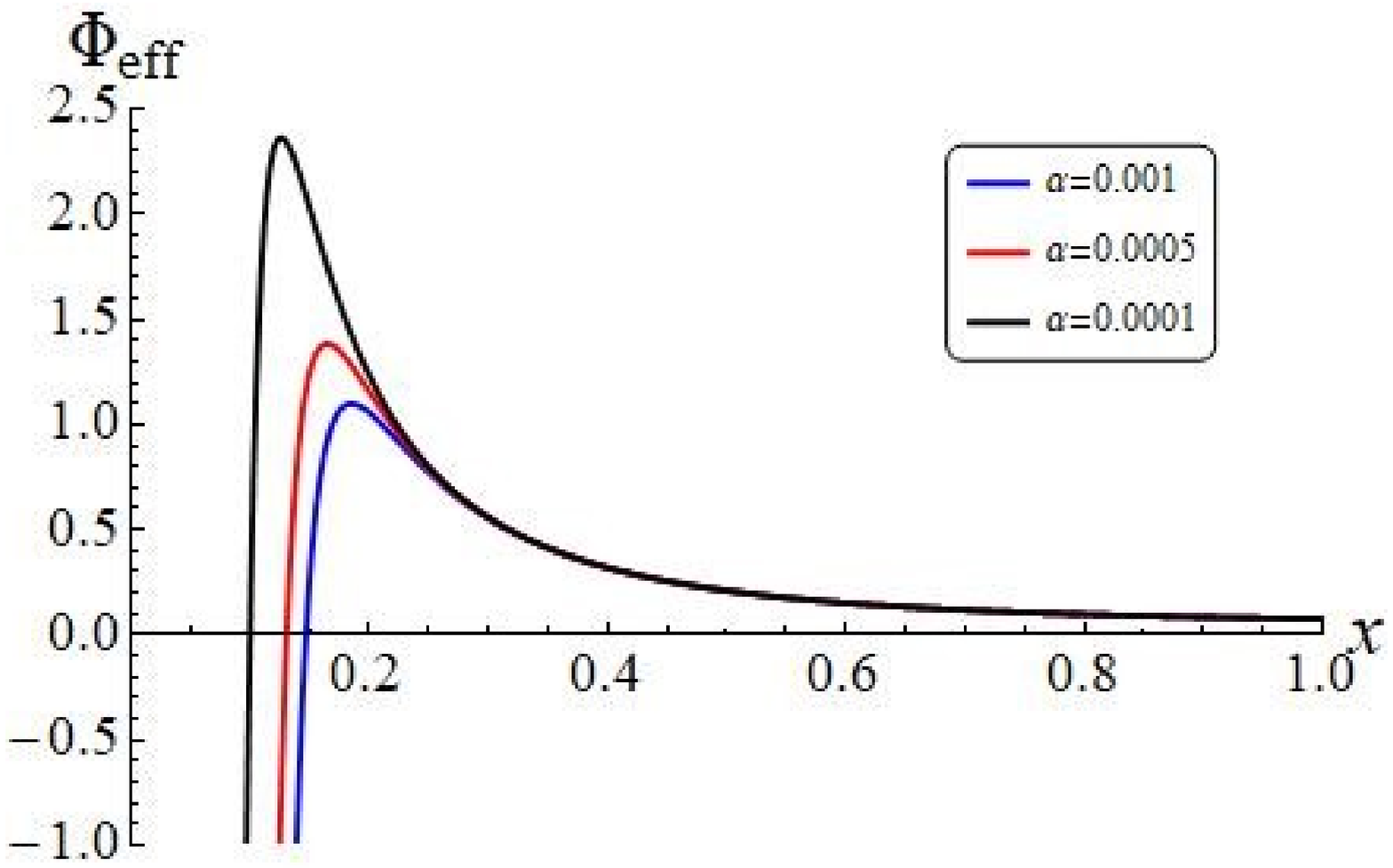}\label{c}}\newline
\caption{(color online).(a): The effective temperature $T_{eff}$ versus $x$
with different nonlinearity parameter $\protect\alpha $ : $\protect\alpha %
=0.001$ (blue line), $\protect\alpha =0.0005$ (red line),$alpha=0.0001$
(black line). (b): The effective pressure $p_{eff}$ versus $x$ with
different nonlinearity parameter $\protect\alpha $ : $\protect\alpha =0.001$
(blue line), $\protect\alpha =0.0005$ (red line),$alpha=0.0001$ (black
line). (c): The effective electric potential $\Phi _{eff}$ versus $x$ with
different nonlinearity parameter $\protect\alpha $ : $\protect\alpha =0.001$
(blue line), $\protect\alpha =0.0005$ (red line), $\protect\alpha =0.0001$
(black line). Here, $q=0.1,n=4$. }
\end{figure}
The first law of thermodynamics is to study the universal relations of the
thermal properties. The state parameters of HDBRN space-time satisfies with
the first thermodynamic relation formula is written as \cite{49,50}
\begin{equation}
dM={T_{eff}}dS-{P_{eff}}dV+{\Phi _{eff}}dQ,  \label{3.1}
\end{equation}%
where the effective temperature $T_{eff}$, effective electric potential $%
\Phi _{eff}$ and the effective pressure $P_{eff}$ are denoted as
\begin{equation}
{T_{eff}}={\left( {\frac{{\partial M}}{{\partial S}}}\right) _{q,V}}=\frac{{{%
{\left( {\frac{{\partial M}}{{\partial x}}}\right) }_{{r_{c}}}}{{\left( {%
\frac{{\partial V}}{{\partial {r_{c}}}}}\right) }_{x}}-{{\left( {\frac{{%
\partial V}}{{\partial x}}}\right) }_{{r_{c}}}}{{\left( {\frac{{\partial M}}{%
{\partial {r_{c}}}}}\right) }_{x}}}}{{{{\left( {\frac{{\partial S}}{{%
\partial x}}}\right) }_{{r_{c}}}}{{\left( {\frac{{\partial V}}{{\partial {%
r_{c}}}}}\right) }_{x}}-{{\left( {\frac{{\partial V}}{{\partial x}}}\right) }%
_{{r_{c}}}}{{\left( {\frac{{\partial S}}{{\partial {r_{c}}}}}\right) }_{x}}}}%
,  \label{3.2}
\end{equation}%
\begin{equation}
{\Phi _{eff}}={\left( {\frac{{\partial M}}{{\partial Q}}}\right) _{S,V}},
\label{3.3}
\end{equation}%
\begin{equation}
{P_{eff}}=-{\left( {\frac{{\partial M}}{{\partial V}}}\right) _{q,S}}=-\frac{%
{{{\left( {\frac{{\partial M}}{{\partial x}}}\right) }_{{r_{c}}}}{{\left( {%
\frac{{\partial S}}{{\partial {r_{c}}}}}\right) }_{x}}-{{\left( {\frac{{%
\partial S}}{{\partial x}}}\right) }_{{r_{c}}}}{{\left( {\frac{{\partial M}}{%
{\partial {r_{c}}}}}\right) }_{x}}}}{{{{\left( {\frac{{\partial V}}{{%
\partial x}}}\right) }_{{r_{c}}}}{{\left( {\frac{{\partial S}}{{\partial {%
r_{c}}}}}\right) }_{x}}-{{\left( {\frac{{\partial S}}{{\partial x}}}\right) }%
_{{r_{c}}}}{{\left( {\frac{{\partial V}}{{\partial {r_{c}}}}}\right) }_{x}}}}%
.  \label{3.4}
\end{equation}%
Here, the thermodynamic volume is that between the black hole horizon and
the cosmological horizon, namely \cite{27}
\begin{equation}
V={V_{c}}-{V_{+}}=\frac{{{V_{n-1}}r_{c}^{n}}}{n}\left( {1-{x^{n}}}\right) .
\label{3.5}
\end{equation}%
Considering the dimension, we assume the space time entropy is
\begin{equation}
S=\frac{{{V_{n-1}}r_{c}^{n-1}}}{4}{F_{n}}(x),  \label{3.6}
\end{equation}%
where ${F_{n}}(x)$ is an arbitrary function of $x$.

Substituting Eq. (\ref{2.11}), Eq. (\ref{3.5}) and Eq. (\ref{3.6}) into Eq. (%
\ref{3.2}), one can obtain
\begin{equation}
{T_{eff}}=\frac{{(n-1)B(x,q)}}{{4\pi {r_{c}}(1-{x^{n}})A(x)}},  \label{3.7}
\end{equation}%
where
\begin{eqnarray}
A(x) &=&{F_{n}}^{\prime }(x)\left( {1-{x^{n}}}\right) +(n-1){F_{n}}(x),{F_{n}%
}(x)=1+{x^{n-1}}+{f_{n}}(x),  \notag \\
B(x,q) &=&k{x^{n-3}}\left( {(n-2-n{x^{2}})(1-{x^{n}})+2(n-1){x^{n}}(1-{x^{2}}%
)}\right)  \notag \\
&&+\frac{{2{q^{2}}}}{{(n-1)(n-2)r_{c}^{2(n-2)}{x^{n-1}}}}\left( {n{x^{n}}(1-{%
x^{n-2}})-(n-2)(1-{x^{3n-2}})}\right)  \label{3.8} \\
&&+\frac{{4{q^{4}}\alpha }}{{(3{n^{2}}-7n+4)r_{c}^{4n-6}{x^{3n-3}}}}\left( {{%
x^{n}}(3n-4)(1-{x^{4n-4}})-(4n-4){x^{n}}+n{x^{4n-4}}+(3n-4)}\right) .  \notag
\end{eqnarray}%
When we set $T_{+}=T_{c}$ , the effective temperature of the spacetime also
has the same value
\begin{equation}
{T_{c}}={T_{+}}=T={T_{eff}}=\frac{{\tilde{B}(x,q)}}{{4\pi {r_{c}}(1-{x^{n}}%
)A(x)}}.  \label{3.9}
\end{equation}%
We can set $\frac{{2{q^{2}}}}{{r_{c}^{2n-4}}}$ as a function of $\frac{{4{%
q^{4}}\alpha }}{{r_{c}^{4n-6}}}$ and $k$ , then substitute Eq. (\ref{2.13})
into Eq. (\ref{3.8}), one obtains
\begin{eqnarray}
\frac{{\tilde{B}(x,q)K(x,n)}}{{2{x^{n-2}}(1+{x^{n+1}})}} &=&k(n-2)(n-1)[n{%
x^{n-2}}(1-{x^{2}})(1-{x^{n-2}})-(1-{x^{n}})(1-{x^{2n-4}})]  \notag \\
&&+\frac{{(1-{x^{2n-2}})}}{{(3{n^{2}}-7n+4){x^{3n-4}}}}\frac{{4{q^{4}}\alpha
}}{{r_{c}^{4n-6}}}[4{x^{n-2}}(1-{x^{n}})+n(1-11{x^{n-2}}+11{x^{2n-2}}-{%
x^{3n-4}})  \notag \\
&&-2{n^{2}}(1-{x^{3n-4}}+5{x^{2n-2}}-5{x^{n-2}})+{n^{3}}(1-{x^{3n-4}}-3{%
x^{n-2}}+3{x^{2n-2}})].  \label{3.10}
\end{eqnarray}%
Substituting Eq. (\ref{2.15})and Eq. (\ref{3.10}) into Eq. (\ref{3.9}),
\begin{equation}
A(x)=\frac{{(n-1){x^{n-2}}(1+{x^{n+1}})}}{{(1-{x^{n}})}}.  \label{3.11}
\end{equation}%
When $\frac{{4{q^{4}}\alpha }}{{r_{c}^{4n-6}}}$ as a function of $\frac{{2{%
q^{2}}}}{{r_{c}^{2n-4}}}$ and $k$ , then substitute Eq. (\ref{2.13}) into
Eq. (\ref{3.8}), we can get
\begin{eqnarray}
&&\frac{{\tilde{B}(x,q)\tilde{K}(x,n)(3{n^{2}}-7n+4)}}{{2{x^{n-2}}(1+{x^{n+1}%
})}}  \notag \\
&=&k(n-1)(n-2)\left[ {4(1-{x^{n}})(1-{x^{4n-6}})}\right. +n[-7(1+{x^{5n-6}}%
)-3{x^{n-2}}(1+{x^{3n-2}})+10{x^{n}}(1+{x^{3n-6}})]  \notag \\
&&+{n^{2}}[5{x^{n-2}}(1+{x^{3n-2}})-8{x^{n}}(1+{x^{3n-6}})+3(1+{x^{5n-6}})]+2%
{n^{3}}[{x^{n}}(1+{x^{3n-6}})\left. {-{x^{n-2}}(1+{x^{3n-2}})]}\right]
\notag \\
&&-\frac{{2{q^{2}}(1-{x^{2n-2}})}}{{r_{c}^{2n-4}{x^{n-2}}}}\left[ {4{x^{n-2}}%
(1-{x^{n}})}\right. +n(1-11{x^{n-2}}+11{x^{2n-2}}-{x^{3n-4}})  \label{3.12}
\\
&&-2{n^{2}}(1-{x^{3n-4}}+5{x^{2n-2}}-5{x^{n-2}})+{n^{3}}\left. {(1-{x^{3n-4}}%
-3{x^{n-2}}+3{x^{2n-2}})}\right] .  \notag
\end{eqnarray}%
Substituting Eq. (\ref{2.18}) and Eq. (\ref{3.12}) into Eq. (\ref{3.9}), we
have
\begin{equation}
A(x)=\frac{{(n-1){x^{n-2}}(1+{x^{n+1}})}}{{(1-{x^{n}})}}.  \label{3.13}
\end{equation}%
\begin{figure}[tbp]
\centering
\subfigure[]{\includegraphics[width=0.4%
\columnwidth,height=1.2in]{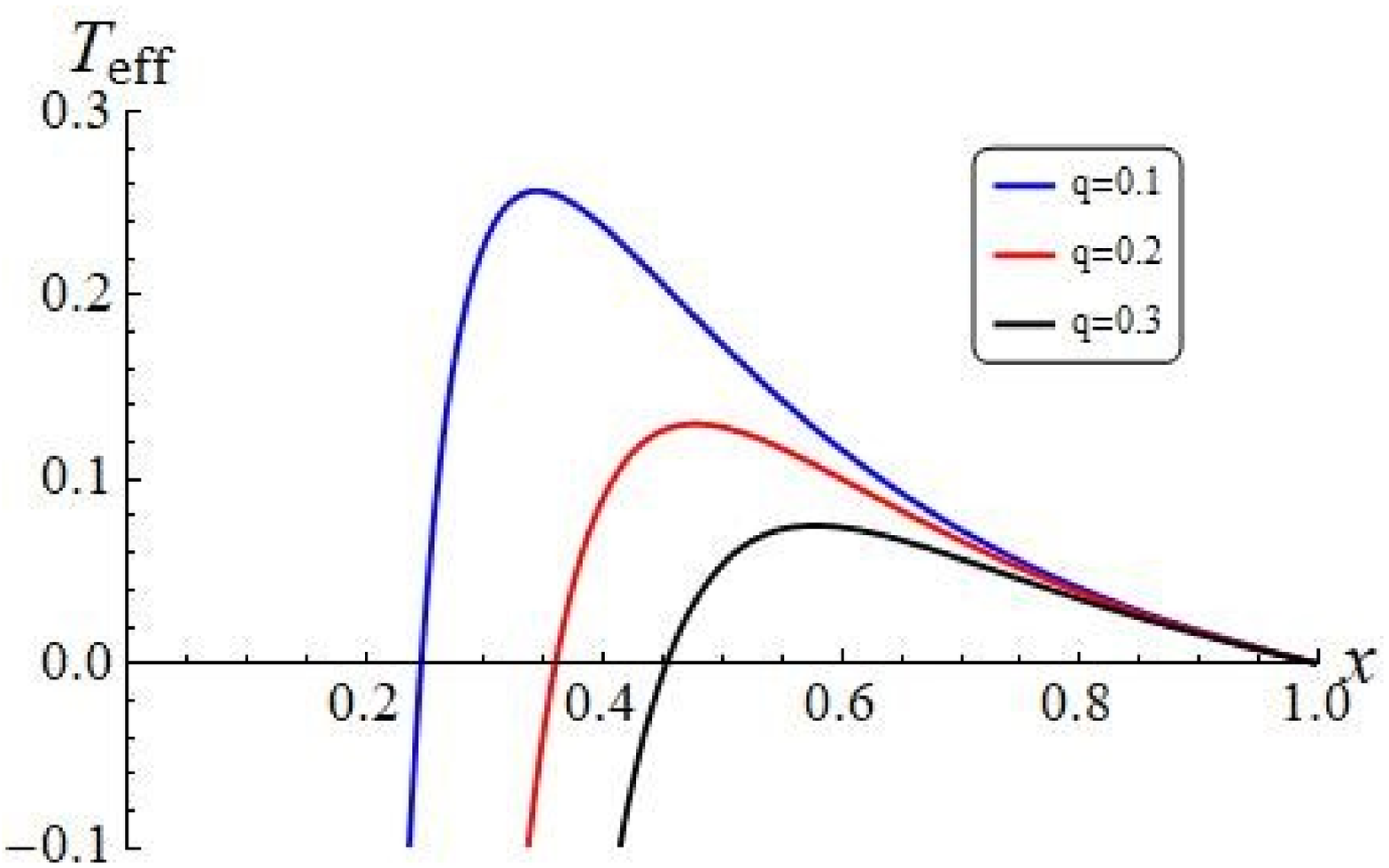}\label{a}}
\subfigure[]{\includegraphics[width=0.4\columnwidth,height=1.2in]{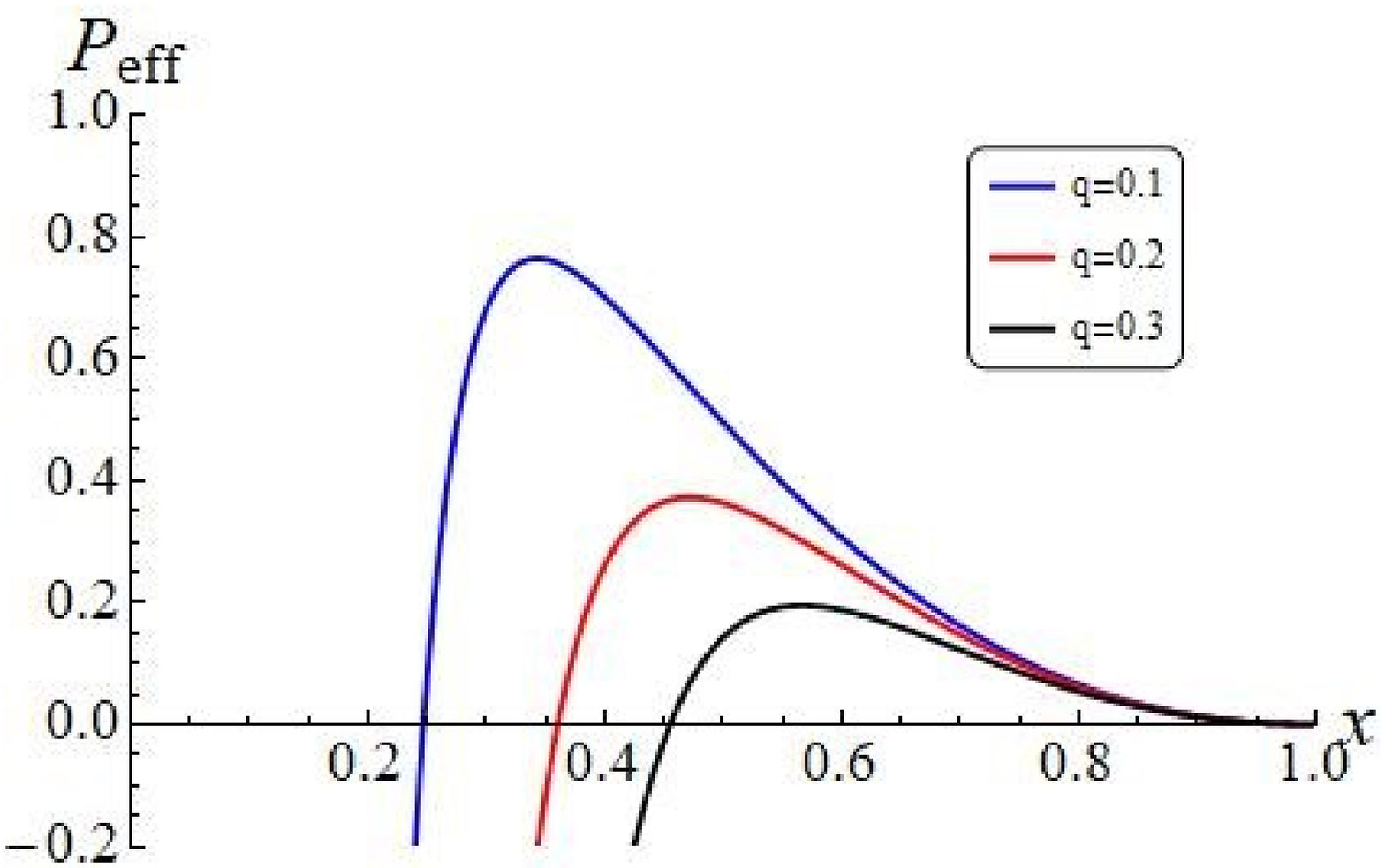}\label{b}
}\subfigure[]{\includegraphics[width=0.4%
\columnwidth,height=1.2in]{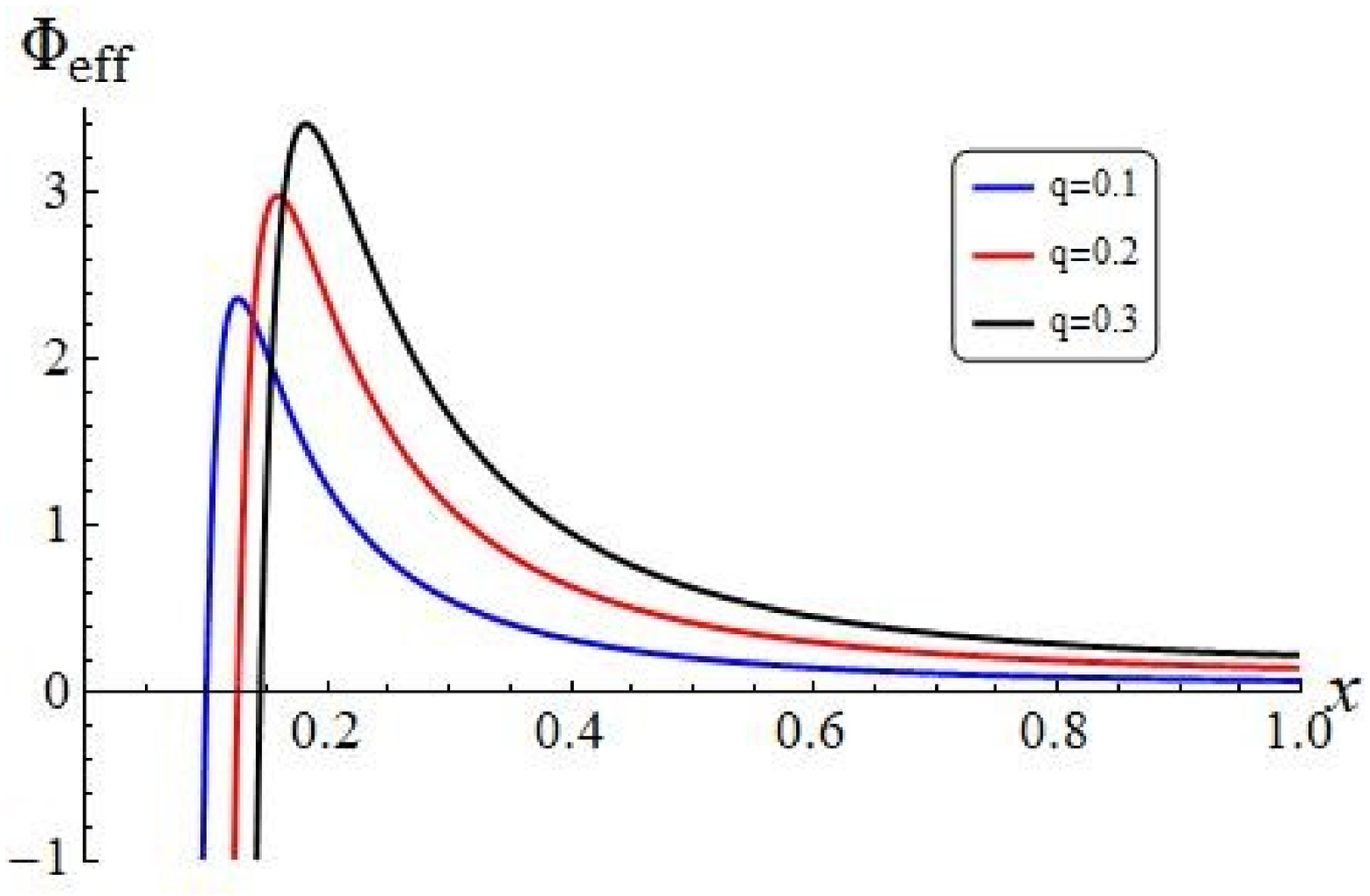}\label{c}}\newline
\caption{(color online).(a): The effective temperature $T_{eff}$ versus $x$
with the different values of $q$ : $q=0.1$ (blue line), $q=0.2$ (red line),$%
q=0.3$ (black line). (b): The effective pressure $p_{eff}$ versus $x$ with
the different values of $q$ : $q=0.1$ (blue line), $q=0.2$ (red line),$q=0.3$
(black line). (c): The effective electric potential $\Phi _{eff}$ versus $x$
with the different values of $q$ : $q=0.1$ (blue line), $q=0.2$ (red line),$%
q=0.3$ (black line). Here, $\protect\alpha =0.0001,n=4$. }
\end{figure}
From Eq. (\ref{3.11}) and Eq. (\ref{3.13}), one can find that the system
entropy satisfied the solution of the differential equation is independent
of the selected variables. The solution of Eq. (\ref{3.13}) can be written
as
\begin{eqnarray}
{F_{n}}(x) &=&\frac{{3n-1}}{{2n-1}}{(1-{x^{n}})^{(n-1)/n}}-\frac{{n(1+{%
x^{2n-1}})-(2n-1){x^{n-1}}(1+x)}}{{(2n-1)(1-{x^{n}})}}  \notag \\
&=&\frac{{3n-1}}{{2n-1}}{(1-{x^{n}})^{(n-1)/n}}-\frac{{n(1+{x^{2n-1}}%
)+(2n-1)(1-2{x^{n}}-{x^{2n-1}})}}{{(2n-1)(1-{x^{n}})}}+1+{x^{n-1}}  \notag \\
&=&{f_{n}}(x)+1+{x^{n-1}.}  \label{3.14}
\end{eqnarray}%
The case of calculating the Eq (\ref{3.13}) is permitted for ${F_{n}}(0)=1$
or ${f_{n}}(0)=0$. It is based on that $x\rightarrow 0$, ${r_{+}}<<{r_{c}}$,
which means the spacetime verge to the pure de Sitter spacetime.

From Eq. (\ref{3.11}) and Eq. (\ref{3.7}), one can obtain the effective
temperature
\begin{equation}
{T_{eff}}=\frac{{B(x,q)}}{{4\pi {r_{c}}{x^{n-2}}(1+{x^{n+1}})}}.
\label{3.15}
\end{equation}%
The effective pressure is
\begin{equation}
{P_{eff}}=\frac{{{{(n-1)}^{2}}g(x,q)}}{{4\pi r_{c}^{2}(1-{x^{n}})A(x)}}=%
\frac{{(n-1)g(x,q)}}{{4\pi r_{c}^{2}{x^{n-2}}(1+{x^{n+1}})}},  \label{3.16}
\end{equation}%
with
\begin{eqnarray}
g(x,n) &=&k\left( {\frac{{{x^{n-3}}(1+{x^{n-1}})(n-2+2{x^{n}}-n{x^{2}})}}{{%
(1-{x^{n}})}}-{x^{n-2}}(n-2)({x^{n-2}}-{x^{n}})}\right) -\frac{{4{q^{4}}%
\alpha }}{{(3{n^{2}}-7n+4)r_{c}^{2(2n-3)}}}  \notag \\
&&\left( {\frac{{(1+{x^{n-1}})}}{{1-{x^{n}}}}\frac{{(4n-4){x^{n}}-n{x^{4n-4}}%
-(3n-4)}}{{{x^{3n-3}}}}-\frac{{(3n-4)(1-{x^{4n-4}})}}{{{x^{2n-2}}}}}\right)
\label{3.17} \\
&&+\frac{{2{q^{2}}}}{{(n-1)(n-2)r_{c}^{2(n-2)}}}\left( {\frac{{(1+{x^{n-1}})}%
}{{1-{x^{n}}}}\frac{{(2n-2){x^{n}}-n{x^{2n-2}}-(n-2)}}{{{x^{n-1}}}}+(n-2)(1-{%
x^{2n-2}})}\right) .  \notag
\end{eqnarray}

The effective electric potential is
\begin{eqnarray}
{\Phi _{eff}} &=&{\left( {\frac{{\partial M}}{{\partial Q}}}\right) _{S,V}}
\notag \\
&=&\frac{{(n-1)(1-{x^{2n-2}})q}}{{(1-{x^{n}})r_{c}^{n-2}{x^{n-2}}}}\left[ {%
\frac{1}{{(n-1)(n-2)}}-\frac{{4{q^{2}}\alpha }}{{(3{n^{2}}-7n+4)r_{c}^{2n-2}}%
}\frac{{(1+{x^{2n-2}})}}{{{x^{2n-2}}}}}\right] .  \label{3.18}
\end{eqnarray}

From Eq. (\ref{3.15}), Eq. (\ref{3.16}) and Eq. (\ref{3.18}), we can plot
the ${T_{eff}}-x,{P_{eff}}-x,{\Phi _{eff}}-x$ curves for the different
values of $n$ in the Fig. 2, respectively. From Fig. 2(a), one can observe
that the region of spacetime effective temperature $T_{eff}$ greater than
zero is inversely proportional to the number of space-time dimensions. We
can also find the maximal value of effective pressure $P_{eff}$is
proportional to the number of space-time dimensions when the value of q and $%
\alpha $ is fixed, as shown in Fig. 2(b). In addition, Fig. 2(c) indicates
that the maximal value of effective electric potential $\phi _{eff}$ is also
proportional to the number of space-time dimensions when the value of $q$
and $\alpha $ is fixed.

Considering Eq. (\ref{3.15}), Eq. (\ref{3.16}) and Eq. (\ref{3.18}), we can
also plot Fig. 3. This figure show the how the effective temperature,
pressure and electric potential versus $x$ with the different values of $%
\alpha $. As shown in Fig. 3(a) and Fig. 3(b), when the values of $n$ and $q$
are fixed, the effective temperature $T_{eff}$ and the effective pressure $%
P_{eff}$ is not effected by $\alpha $. From Fig. 3(c), one can find that the
maximal value of effective electric potential $\Phi _{eff}$ is in direct
proportion to the value of $\alpha $.

In addition, as the same method above mentioned, we also plot the curves
about ${T_{eff}} - x,{P_{eff}} - x,{\Phi _{eff}} - x$ with the different
values of $q$. The Fig.4 indicates that the maximal values of effective
temperature $T_{eff}$ and the effective pressure $p_{eff}$ are inversely
proportional to the value of is inversely direct proportion to the values of
$q$ , while the electric potential is in direct proportion to it.

\section{conclusion and discussion}

People usually think that black hole thermodynamics is one of the key points
to connect gravity and quantum mechanics, and very useful for the study of
quantum gravity theory. There are systematic approaches and rich features
for the research of black hole thermodynamics in AdS space, especially the
criticality, while it's useless for the research of de Sitter spacetime. The
main reason is there are black hole and the cosmological horizons which have
different radiation temperature in general, so the spacetime does not
satisfy the requirement of stability of thermodynamic equilibrium. Secondly
the corresponding thermodynamic systems of two horizons are not independent,
there are some relations between them. Therefore, it is important to find a
thermodynamic system to describe the de Sitter spacetime completely. The
recent references \cite{49,50} proposed a new approach to study the
thermodynamics of dS black holes. We use this approach to the
higher-dimensional de Sitter spacetime with nonlinear electric field and try
to provide the theoretical basis to find the general entropy of the dS black
holes and the effective thermodynamical quantities.

From the discussion in Sect. 3, it is known that to find the entropy of the
dS black holes and the effective thermodynamical quantities, we need to
consider the first law of thermodynamics and the dimension, thus we have
equation (\ref{3.6}). When the radiation temperature of black hole horizon $%
T_{+}$ equals the radiation temperature of the cosmological horizon $T_{c}$,
i.e. $T={T_{+}}={T_{c}}$, then the effective temperature ${T_{eff}}$ should
be same with these two temperature of horizons, i.e. $T={T_{eff}}$. We
obtain the differential equation of spacetime entropy, see Eq. \ref{3.13}).
And from Eq. (\ref{3.11}) and Eq. (\ref{3.13}), we know that the
differential equation has no relation with the independently parameters we
choose. This is same with the research in multi-variables general
thermodynamics. It means that the conclusions from HDBRN spacetime
thermodynamics are self-consistent. When $x\rightarrow 0$, the spacetime
approaches to pure de Sitter as a initial condition and solve the
differential equation (\ref{3.11}) or (\ref{3.13}), then we get the formula
of entropy function (\ref{3.14}). This formula implies that the entropy of
HDBRN spacetime does not include the nonlinear parameters $\alpha $ and $q$,
and it is just the function of the position of spacetime horizon. This is
consistent with the entropy of black hole horizon $S_{+}$ and the entropy of
the cosmological horizon $S_{c}$. What's more, the effective temperature of
HDBRN spacetime given by Eq. (\ref{3.15}) is the function of spacetime
dimension, the position of horizon and the charge of spacetime $q$ and $%
\alpha $, the curves of ${T_{eff}}-x$ reflects the relations between various
parameters. The effective temperature $T_{eff}$ and the pressure $P_{eff}$
of spacetime are independent of the value of $\alpha $ , The maximum value
of the space-time effective potential $\Phi _{eff}$ is proportional to $%
\alpha $ .This may provide the theoretial foundation for the further
research of the classical and quantum features of de Sitter, and its
evolution.

\section*{Acknowledgements}

This work was supported by the National Natural Science Foundation of China
(Grant No.11475108).

\end{document}